\documentclass[11pt]{article}
\usepackage{graphicx}
\usepackage{epsf}

\newcommand{\BaBarPubYear}       {03}

\newcommand{\BaBarConfNumber}    {007}
\newcommand{\SLACPubNumber} {10054}

\def\Journal#1#2#3#4{{#1} {\bf #2}, #3 (#4)}

\def\NIMA{{\em Nucl. Instrum. Methods} A}

\def\PLB{{\em Phys. Lett.}  B}
\def\PRL{\em Phys. Rev. Lett.}
\def\PRD{{\em Phys. Rev.} D}

\input pubboard/babarsym
\def\bmunu     {\ensuremath{\Bp \to \mup \num}\xspace}
\def\btaunu     {\ensuremath{\Bp \to \taup \nut}\xspace}
\def\blnu     {\ensuremath{\Bp \to \ellp \nul}\xspace}

\long\def\inst#1{\par\nobreak\kern 4pt\nobreak
    {\it #1}\par\vskip 10pt plus 3pt minus 3pt}

\begin{document}
{\pagestyle{empty}

\begin{flushright}
\babar-CONF-\BaBarPubYear/\BaBarConfNumber \\
SLAC-PUB-\SLACPubNumber \\
July, 2003
\end{flushright}

\par\vskip 5cm 

% Title of paper

% Title of the paper
\begin{center}
\Large \bf A search for \boldmath{\bmunu}
\end{center}
\bigskip

\begin{center}
\large The \babar\ Collaboration\\
\mbox{ }\\
\today
\end{center}
\bigskip \bigskip

% Abstract
\begin{center}
\large \bf Abstract
\end{center}
We have performed a search for the leptonic decay \bmunu with data 
collected at the \FourS resonance by the \babar\ experiment at the \pep2 storage ring.
With an integrated luminosity of approximately 81.4 \invfb (88.4 million \BB pairs), 
we find no convincing evidence for a signal and set a preliminary upper limit on the branching fraction of
$\BR(\bmunu) < 6.6\times 10^{-6}$ at the 90\% confidence level.  

\vfill
\begin{center}
Presented at the 
International Europhysics Conference On High-Energy Physics (HEP 2003),
7/17---7/23/2003, Aachen, Germany
\end{center}

\vspace{1.0cm}
\begin{center}
{\em Stanford Linear Accelerator Center, Stanford University, 
Stanford, CA 94309} \\ \vspace{0.1cm}\hrule\vspace{0.1cm}
Work supported in part by Department of Energy contract DE-AC03-76SF00515.
\end{center}

\newpage
} % end of pagestyle{empty}

% Input author list file
\begin{center}
\small

The \babar\ Collaboration,
\bigskip

%% author list as of 02-Jun-2003 (595 authors)
%
B.~Aubert,
R.~Barate,
D.~Boutigny,
J.-M.~Gaillard,
A.~Hicheur,
Y.~Karyotakis,
J.~P.~Lees,
P.~Robbe,
V.~Tisserand,
A.~Zghiche
\inst{Laboratoire de Physique des Particules, F-74941 Annecy-le-Vieux, France }
% ...

\end{center}\newpage

% The body of the paper starts here

% reset footnote counter
\setcounter{footnote}{0}

%%%%%%%%%%%%%%%%%%%%%%%%%%%
\section{INTRODUCTION}
\label{sec:Introduction}
%%%%%%%%%%%%%%%%%%%%%%%%%%%

\begin{figure}[htb]
 \centering
 \epsfxsize=4.0in \leavevmode
 \epsffile{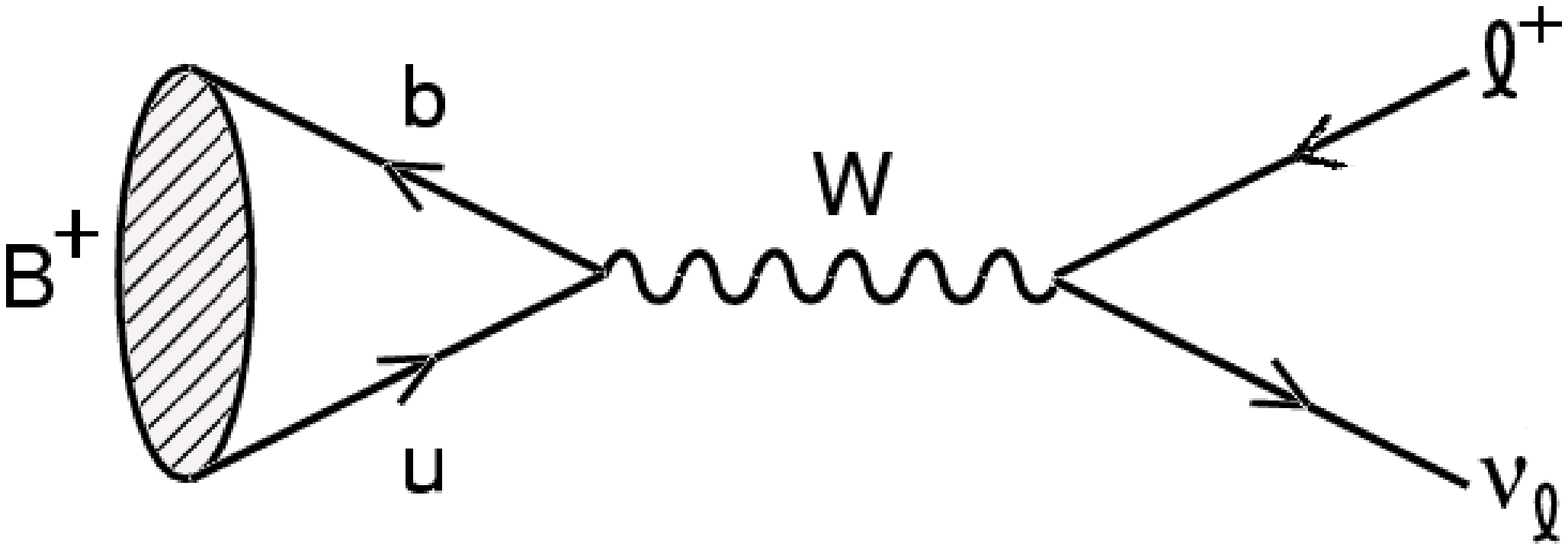}
 \caption{ The SM \blnu annihilation diagram. }
 \label{fig:BtomunuDiagram} 
\end{figure}

The purely leptonic decays \blnu ( $\ell$ = \electron, \mmu, or \mtau )
proceed through the annihilation (Fig.~\ref{fig:BtomunuDiagram}) of the quark-antiquark pair in the 
meson to form a virtual \W boson. In the Standard Model(SM), the branching fraction can be 
calculated as~\footnote{Charge-conjugation is implied
throughout this paper.},    
\begin{equation}
 \BR(\blnu) = \frac{G_{F}^{2} m_{B} m_{\ell}^{2}} {8\pi} 
 \biggl( 1- \frac{m_{\ell}^{2}}{m_{\B}^{2}} \biggr)^{2} \fsubb^{2} |V_{ub}|^{2} 
 \tau_{\B},
\end{equation}
where $G_F$ is the Fermi coupling constant, $m_{\ell}$ and $m_B$ are 
the lepton and \B meson masses, and $\tau_{\B}$ is the \Bp lifetime. The decay rate 
is sensitive to both the Cabibbo-Kobayashi-Maskawa 
matrix element $V_{ub}$ and the \B decay constant \fsubb which describes 
the overlap of the quark wave functions within the meson. Once \Vub has been more precisely 
measured in \B semi-leptonic decays, 
\fsubb could be extracted from  a measurement of the \blnu branching fraction. Currently, 
the uncertainty on \fsubb is one of the main
factors limiting the determination of \Vtd from precision \BzBzb
mixing measurements. 

The Standard Model estimate of the branching fraction for 
\btaunu is about $9\times 10^{-5}$ assuming 
$\tau_{\B}$ = 1.674 \ps, \Vub = 0.0036 and \fsubb = 198 \mev~\cite{PDG}. 
Due to helicity suppression, the expected branching fraction for \bmunu is reduced to 
roughly $4\times 10^{-7}$. However, physics 
beyond the Standard Model could substantially increase these predictions. Charged Higgs boson 
effects may greatly enhance the branching fraction in certain two Higgs doublet models~\cite{Hou}. Similarly, this decay
may be enhanced through mediation by leptoquarks in the Pati-Salam model of quark-lepton
unification~\cite{Valencia}.   

Purely leptonic \B decays have not yet been observed experimentally. The CLEO~\cite{Cleo} and Belle 
Collaborations~\cite{Belle} have set 90\% confidence level upper limits on the \bmunu branching fraction. 
The most stringent limit is currently from Belle,
\begin{eqnarray}
  \BR(\bmunu) < 6.8 \times 10^{-6}\nonumber 
\end{eqnarray}
using 60 \invfb of data collected at the \FourS resonance.

%%%%%%%%%%%%%%%%%%%%%%%%%%%%%%%%%%%%%%%%%%%
\section{THE \babar\ DETECTOR AND DATASET}
\label{sec:babar}
%%%%%%%%%%%%%%%%%%%%%%%%%%%%%%%%%%%%%%%%%%%

The data used in this analysis were collected with the \babar\ detector
at the \pep2 storage ring. The sample consists of an integrated luminosity
of 81.4 \invfb accumulated at the \FourS resonance and 9.6 \invfb accumulated 
at a center-of-mass (CM) energy about 40 \mev below the \FourS resonance.
The on-resonance sample corresponds to about 88.4 million \BB pairs. The collider
is operated with asymmetric beam energies, producing a boost of the \FourS along the 
collision axis of $\beta\gamma = 0.55$ in the laboratory frame. 

The \babar\ detector is optimized for the asymmetric-energy beams at \pep2 and 
is described in detail elsewhere~\cite{babar}. Charged particle momentum and 
direction are measured with a 5-layer double-sided silicon vertex tracker (SVT) and
a 40-layer drift chamber (DCH) which are contained in the 1.5 T magnetic field of a 
superconducting solenoid. Located just outside the DCH, a detector of
internally reflected Cherenkov radiation (DIRC) provides separation of \Kp and \pip. 
The energies of neutral particles 
are measured by the electromagnetic calorimeter (EMC) which is constructed of 6580 CsI(Tl) 
crystals. The flux return of the solenoid is instrumented with resistive plate chambers (IFR) 
for the identification of muons and \KL. Averaged over the momentum and polar angle distributions
of muons from \bmunu, the muon identification efficiency is about 61\%
with a pion misidentification probability of about 2\%.

 A Geant4 based Monte Carlo (MC) simulation was used to optimize the signal 
selection criteria. A sample of 52,000 simulated \BpBm events where
\bmunu and the \Bm decayed generically has been studied. 
Background sources
considered include \epem\to\BB, \epem\to\qqbar 
(\q = \u, \d, \s, and \c), and \epem\to\tautau in quantities comparable 
to the data luminosity. 

%%%%%%%%%%%%%%%%%%%%%%%%%%%
\section{ANALYSIS METHOD}
\label{sec:Analysis}
%%%%%%%%%%%%%%%%%%%%%%%%%%%

 \bmunu is a two-body decay so the 
muon must be mono-energetic in the \B rest frame. The momentum of
the muon is given by
\begin{equation}
 p^* = \frac{m^2_B - m^2_{\mu}}{2m_B} \approx \frac{m_B}{2}.
\end{equation}

\begin{figure} 
 \centering
 \epsfxsize=6.5in \leavevmode
 \epsffile[0 260 568 568]{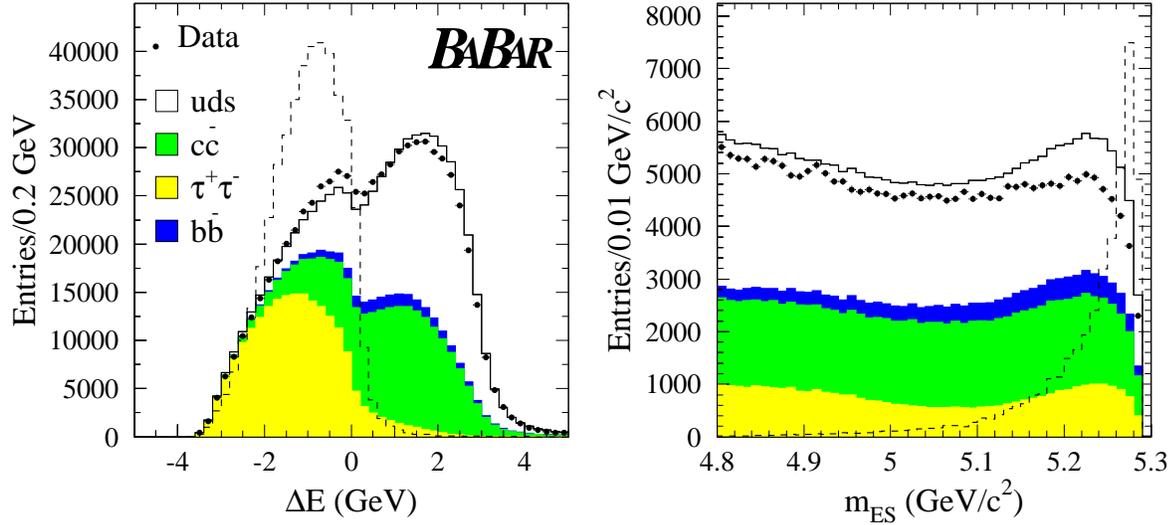}
 \caption{ The distributions of \DeltaE and \mes for on-peak data and MC after muon candidate selection. 
           The signal distributions are overlaid (dashed histograms) with an arbitrary normalization.}
 \label{fig:sigbox_datamc} 
\end{figure}

At \babar, the CM frame is a good approximation to the \B 
rest frame so we initially select well-identified muon candidates with
momentum $p_{CM}$ between 2.25 and 2.95 \gevc in the CM frame. Since the neutrino produced in the 
signal decay is not detected, any other charged tracks or neutral energy in a signal 
event must have been produced by the decay of the companion \B. Therefore, the companion 
\B can be reconstructed from the remaining visible energy in the event.
Signal decays can then be selected using the kinematic variables \DeltaE and energy-substituted mass, \mes, 
defined by
\begin{equation}
 \DeltaE = E_{B}-E_{\rm beam},
\end{equation}
and
\begin{equation}
 \mes = \sqrt{E_{\rm beam}^{2}-|\vec{p}_B|^{\;2}},
\end{equation}
where $\vec{p}_B$ and $E_B$ are the momentum and energy of the reconstructed 
companion \B candidate in the CM frame and $E_{\rm beam}$ is the beam energy in the CM frame. The 
reconstruction of the companion \B includes all 
charged tracks whose distance of closest approach to the beam-spot is less than 1.5 \cm in the $xy$ plane and within 10 \cm along the beam axis.
We also include all neutral calorimeter clusters with 
cluster energy greater than 30 \mev. Particle identification is applied to the charged tracks 
to identify electrons, muons, kaons and protons in order to apply the appropriate mass 
hypothesis to each track and thus improve the \DeltaE and \mes resolution. In addition,
events with additional identified leptons are discarded to discriminate against events containing additional 
neutrinos. The MC indicates that this requirement removes about 24\% of the signal decays. 
Figure~\ref{fig:sigbox_datamc} shows the distributions of \DeltaE and \mes for the 
on-peak data, background MC and signal MC after muon candidate selection. For a signal decay, we expect 
the energy of the companion \B to be consistent with the beam energy in the 
CM frame so that \DeltaE peaks near 0. Due to energy losses from detector acceptance, neutral hadrons and 
additional neutrinos, this distribution is shifted toward negative \DeltaE. We expect that \mes 
should peak near the \B mass for signal decays.

\begin{figure}[!htb]
\begin{center}
\epsfxsize=3.5in \leavevmode
\epsffile[0 0 460 460]{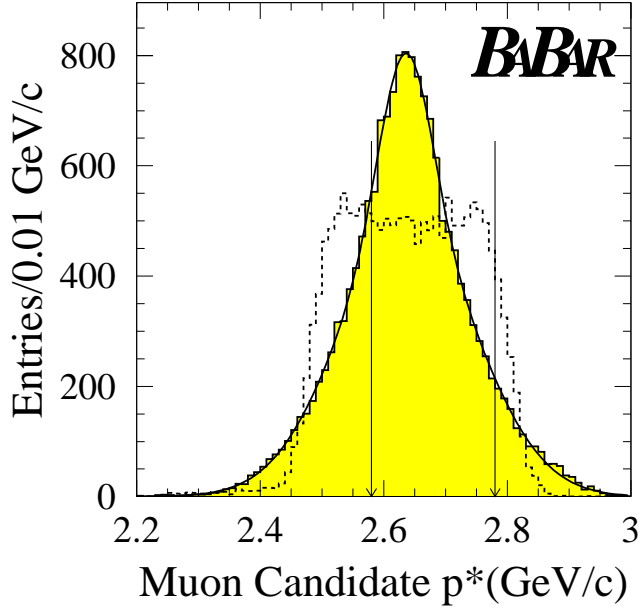}
\caption{ The muon candidate momentum distribution in the reconstructed \B rest frame for all muon candidates
          in the signal MC. The dashed curve is the momentum distribution of the same events in the CM frame.
          The arrows indicate the selected signal region. }
\label{fig:trkprest}
\end{center}
\end{figure}

Once the companion \B is reconstructed, we make a refined estimate of the muon momentum in the \B ``rest'' 
frame. We use the momentum direction of the companion \B and assume a total momentum of 320 \mevc in the CM frame (from the decay 
of the \FourS\to\BB) 
to boost the muon candidate into the reconstructed \B rest frame. Figure~\ref{fig:trkprest} shows the muon candidate 
momentum distribution in the \B rest frame, $p^*$, for all muon candidates in the signal MC. The dashed curve is 
the momentum distribution of the same events in the CM frame. 

Backgrounds may arise from any process producing charged tracks in the 
momentum range of the signal, particularly if the charged tracks are true 
muons. The two most significant backgrounds are the \B semi-leptonic decays involving
$b\rightarrow u\mu\nu$ transitions where the endpoint of the muon 
spectrum approaches that of the signal, and non-resonant \qqbar (continuum) events 
where a charged pion 
is mistakenly identified as a muon. In the continuum events, there must also be 
significant missing energy due to detector acceptance, neutral hadrons, or additional neutrinos 
that mimic the signature of the expected neutrino. We are able to reduce these backgrounds by tightening the selection on 
the muon candidate momentum to 2.58 $< p^* <$ 2.78 \gevc. The cut is asymmetric about the signal peak
due to the decreasing momentum distribution of the backgrounds.  

\begin{figure} 
 \centering
 \epsfxsize=6.5in \leavevmode
 \epsffile[0 260 568 568]{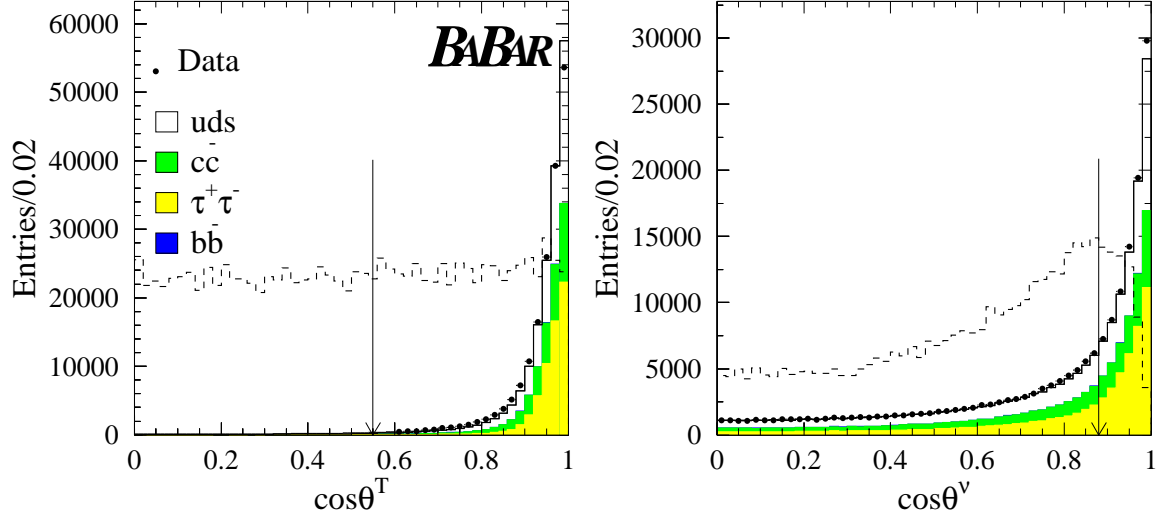}
 \caption{ The distributions of $|\cos\theta^T|$ and $|\cos\theta^{\nu}|$ for on-peak data and MC. The events in these 
           plots have passed the requirement $2.58 < p* <2.78$ \gevc.
           The signal distributions are overlaid (dashed histograms) 
           with an arbitrary normalization.}
 \label{fig:cont_rejection_datamc} 
\end{figure}

Continuum backgrounds are further suppressed using event shape variables. The light-quark events
tend to produce a jet-like event topology as opposed to \BB events which tend to be 
more spherical. We define a variable, $\theta^T$, which is the angle between the 
muon candidate momentum and the thrust axis of the rest of the event in the CM frame. For continuum background,
$|\cos\theta^T|$ peaks sharply near one while the distribution is nearly flat for signal decays. By requiring $|\cos\theta^T| < 0.55$,
we are able to remove about 98\% of the continuum background while retaining about 53\% of the signal decays.
We also use the direction of the missing momentum in the laboratory frame to discriminate against continuum backgrounds.
The missing 4-momentum is calculated as
\begin{equation}
  P^{\nu} = P^{\FourS } - P^B - P^{\mu},
\end{equation}  
where $P^{\FourS }$ is determined from the beam energies, and 
$P^B$ and $P^{\mu}$ represent the reconstructed companion \B and 
signal muon candidate respectively. In continuum decays, 
the missing momentum is often due to undetected particles that were outside the detector acceptance so that 
the polar angle of the missing momentum, $\theta^{\nu}$, lies near the beam axis. Therefore, we 
require $|\cos\theta^{\nu}| <$ 0.88 so that 
the missing momentum is directed into the detector's fiducial volume. In addition, we require that the event contains 
at least four charged tracks and that the normalized second Fox-Wolfram moment~\cite{FoxWolf} is less than 0.98.  
Figure~\ref{fig:cont_rejection_datamc} shows the on-peak data and MC distributions of 
$|\cos\theta^T|$ and $|\cos\theta^{\nu}|$. The events in these plots have passed the requirement $2.58< p* <2.78$ \gevc. 
For comparison, the signal MC is overlaid with an arbitrary normalization. The remaining \BB background is not visible due 
to the earlier requirement on $p^*$. 

We select \bmunu signal events with simultaneous requirements on \DeltaE and \mes, thus 
forming a ``signal box.'' The dimensions of the signal box, as well as the 
above requirements on $p^*$, $|\cos\theta^T|$ and $|\cos\theta^{\nu}|$, were determined using an optimization 
procedure that searches the cut-parameter space to find the combination of cuts that maximizes the quantity 
$S/\sqrt{S+B}$ where $S$ and $B$ are the expected signal and background yields in the MC respectively. 
The signal branching fraction was set to 3$\times 10^{-7}$ during the optimization procedure. 
The resulting box is defined by 
-0.75 $<$ \DeltaE $<$ 0.5 \gev and \mes $>$ 5.27 \gevcc. The MC suggests that about 25\% of signal decays 
passing all previous cuts fall within the signal box. After applying all selection criteria, the 
\bmunu efficiency is determined from the MC to be 2.24 $\pm$ 0.07 \% where the uncertainty is due 
to MC statistics. 
Figure~\ref{fig:unblind}(a) shows the distribution of \DeltaE vs \mes for the signal MC. The signal box 
is represented by the solid lines.

In addition to the signal box, we have defined a slightly larger blinding box. The data within the blinding 
box was kept hidden until the analysis was completed in order to remove any possible experimenter's bias. 
Finally, we define several 
sideband regions that will be denoted the fit sideband and the \DeltaE sidebands. The boundaries of these regions 
in the (\DeltaE, \mes) plane are listed in table~\ref{tab:boxes}. 

\begin{table}[!htb]
 \caption{ The boundaries of the signal box and various sidebands defined for this analysis. }
 \begin{center}
 \begin{tabular}{|l|c|c|} \hline 
   region                       & \DeltaE range (\gev) & \mes range (\gevcc)  \\
\hline
   signal box(SB)               & [ -0.75, 0.50 ]             & $>$ 5.27 \\
   blinding box                 & [ -1.30, 0.70 ]             & $>$ 5.24 \\  
   fit sideband                 & [ -0.75, 0.50 ]             & [ 5.10, 5.24 ] \\
   \DeltaE sideband (bottom)    & [ -3.00, -1.30 ]            & $>$ 5.10 \\
   \DeltaE sideband (top)       & [ 0.70, 1.50 ]              & $>$ 5.10 \\
\hline 
 \end{tabular}
 \end{center}
 \label{tab:boxes}
\end{table}

\begin{figure}[hbt]
 \centering
 \epsfxsize=4.0in \leavevmode
 \epsffile[0 0 470 470]{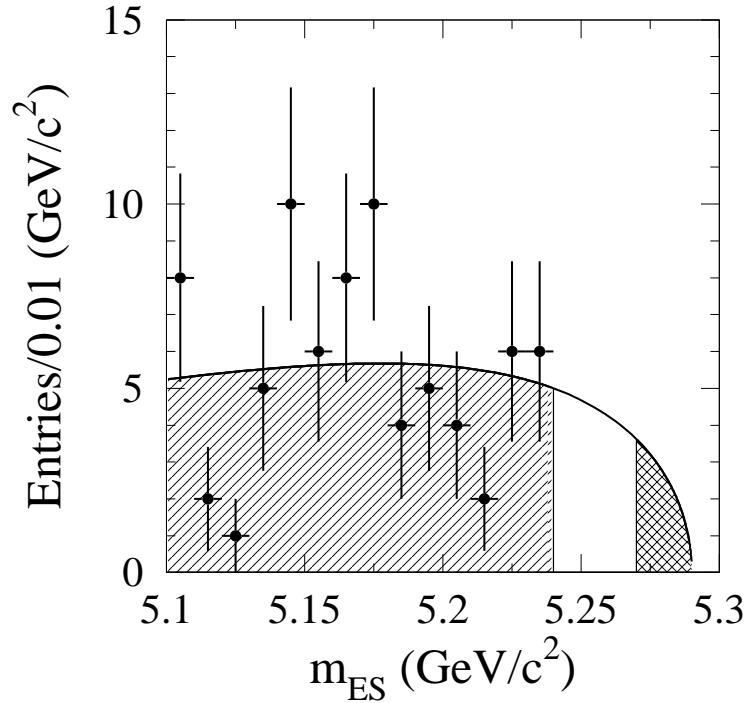}
 \caption{ Results of the Argus fit for the background estimate. The fit is performed only on the region
          5.10 $<$ \mes $<$ 5.24 \gevcc (diagonally shaded area) and extrapolated into the signal region represented by the 
          crosshatched area.}
 \label{fig:fit} 
\end{figure}

We estimate the background in the signal box assuming that the \mes distribution is described by the Argus function~\cite{Argus}. 
This assumption is consistent with the observed distributions in the MC and data \DeltaE sidebands. Due to the correlation between \DeltaE
and \mes introduced by the inclusive reconstruction of the companion \B, it is not possible to determine the Argus shape from the \DeltaE 
sidebands. Therefore, the Argus parameter is determined from an unbinned maximum likelihood fit using the data in the region 
defined by -0.75 $<$ \DeltaE $<$ 0.5 \gev and 5.10 $<$ \mes $<$ 5.24 \gevcc. The Argus shape ($A$) is extrapolated 
through the signal box and constrained to be 0 at the endpoint which is fixed at 5.29 \gevcc. Figure~\ref{fig:fit} shows the 
results of the fit. The expected background is given by,
\begin{equation}  
  N_{\rm bkg} = N_{\rm fit} \times \frac{\int^{5.29}_{5.27}A(\mes)d\mes}{\int^{5.24}_{5.10}A(\mes)d\mes} \equiv N_{\rm fit}\times R_{\rm Argus}
  \label{eq:new_extrap}
\end{equation}
where $N_{\rm fit}$ is the number of events contributing to the fit. The ratio $R_{\rm Argus}$ is the area of the background probability density 
function in the signal box divided by the area within the fit region. The result is $N_{\rm bkg} = 5.0^{+1.8}_{-1.4}$
events. The uncertainty is determined by varying the Argus parameter by the $\pm 1\sigma$ uncertainty from the fit. This technique may 
underestimate the background since any peaking component within the blinding box would not 
be accounted for. Therefore, any upper limit 
obtained on the branching fraction should be conservative. When this 
procedure is applied 
to the MC, the resulting background estimate is $5.2 \pm 0.5$ events, in agreement with the true value of $5.7 \pm 0.5$ events. 
The MC indicates the background in the signal box after all cuts are applied is composed of 
about 58\% \qqbar (where $q$ = $u$, $d$, or $s$), 22\% \ccbar and 20\% \BB events.

%%%%%%%%%%%%%%%%%%%%%%%%%%%%%%%%%%
\section{SYSTEMATIC STUDIES}
\label{sec:Systematics}
%%%%%%%%%%%%%%%%%%%%%%%%%%%%%%%%%%

To set an upper limit on the \bmunu branching fraction we must evaluate systematic 
uncertainties in the luminosity (number of \Bpm in the sample), the background estimate, and the signal 
efficiency. The number of \Bpm mesons in the on-peak data sample is estimated to be 88.4 million with an uncertainty of 
1.1\%. The uncertainty in the background estimate is determined by varying the Argus shape within the 
$\pm 1\sigma$ uncertainty from the fit.

The uncertainty in the signal efficiency includes the muon candidate selection (particle identification and 
tracking efficiency) 
as well as the reconstruction efficiency of the companion \B. The muon identification efficiency has been
studied using muon control samples taken from \epem\to\epem\mumu events in the data. For muons in the 
momentum and polar angle region of our signal candidates, we estimate the uncertainty in the identification efficiency 
from the control sample to be 4.2\%. The tracking efficiency for the muon candidate was evaluated from the fraction 
of tracks reconstructed in the SVT that are also found in the DCH. We find that the tracking efficiency is overestimated 
in the MC by 0.8\%. Therefore, we reduce the MC signal efficiency by this amount and assume a systematic error of 2\%.  

The companion \B reconstruction has been studied using a control sample of 
$B^+\rightarrow D^{(*)0}\pi^+$ events. This is also a two-body decay so it is topologically very similar to our 
signal. Once reconstructed, the pion can be treated as if it were the signal muon and the $D^{(*)0}$ decay products 
can be ignored to simulate the neutrino. Then the companion \B is reconstructed in the control sample as it would be for
signal. We then compare the efficiencies for each of our companion \B selection cuts in the $B^+\rightarrow D^{(*)0}\pi^+$
data and MC to quantify any data/MC discrepancies that may affect the signal efficiency. We find that the efficiency 
after all selection cuts is lower in the data by a factor of 0.94$\pm$0.04 where the uncertainty is due to the statistics 
of the data and MC control samples. The signal efficiency obtained from the MC is therefore corrected by this factor. A 
summary of the systematic uncertainties in the signal efficiency is given in table~\ref{tab:systematics}. After applying the 
necessary corrections to the MC, we estimate the signal efficiency to be 2.09 $\pm$ 0.06 \stat $\pm$ 0.13 \syst \%.    

\begin{table}[!htb]
 \caption{ Contributions to the systematic uncertainty on the signal efficiency. }
 \begin{center}
 \begin{tabular}{|l|c|c|} \hline 
   source                       & correction & uncertainty   \\
\hline
   tracking efficiency          & 0.992        & 2.0\%         \\
   muon identification          & -            & 4.2\%         \\
   companion \B reconstruction  & 0.94         & 4.3\%         \\
\hline
   total                        & 0.932        & 6.3\%         \\
\hline 
 \end{tabular}
 \end{center}
 \label{tab:systematics}
\end{table} 

%%%%%%%%%%%%%%%%%%%%%%%%%%%
\section{PHYSICS RESULTS}
\label{sec:Results}
%%%%%%%%%%%%%%%%%%%%%%%%%%%

\begin{figure} 
 \centering
 \epsfxsize=6.5in \leavevmode
 \epsffile[0 265 530 530]{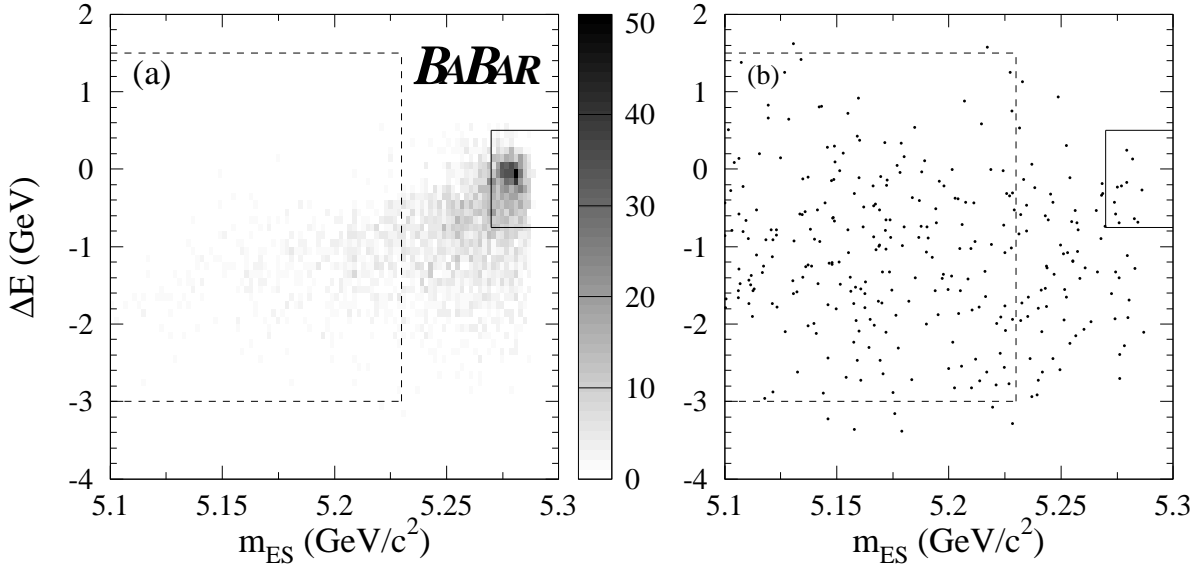}
 \caption{ The distributions of \DeltaE vs \mes in the \bmunu signal MC (a) and
           the on-resonance data (b). The signal box is represented by the solid line while the grand 
           sideband is represented by the dashed line.}
 \label{fig:unblind} 
\end{figure}

\begin{figure} 
 \centering
 \epsfxsize=5.0in \leavevmode
 \epsffile[15 30 470 470]{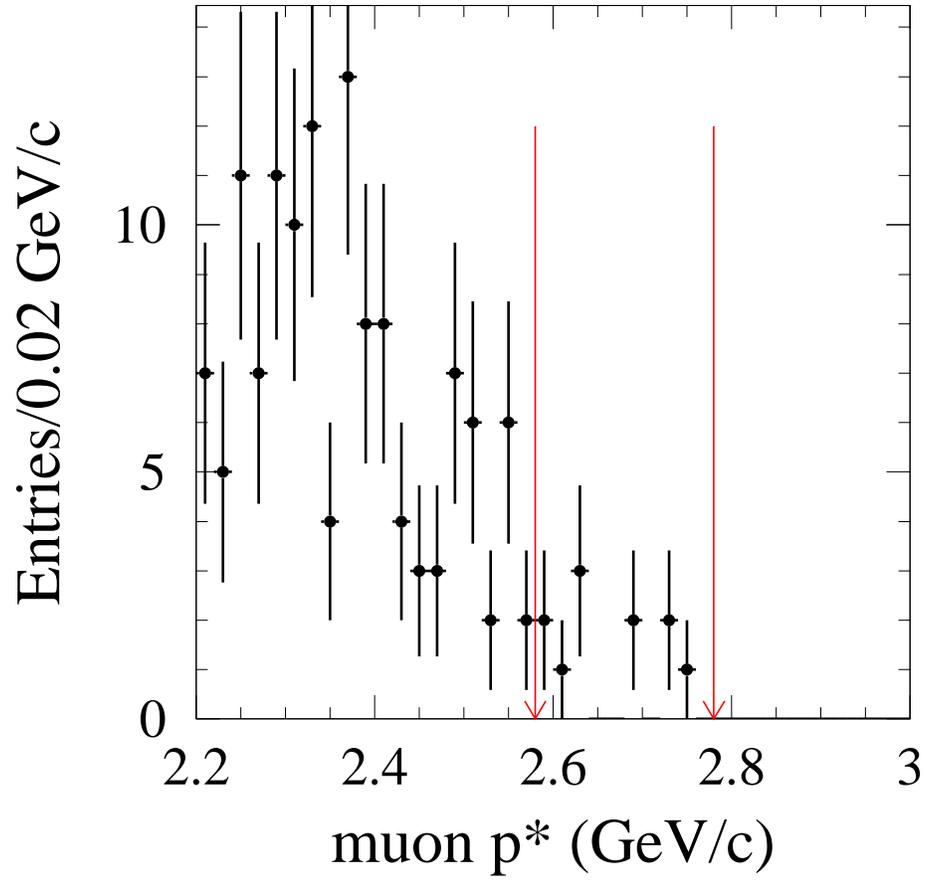}
 \caption{ The $p*$ distribution of the on-resonance data after all other selection 
           cuts have been applied. The region between the two arrows ($2.58 < p* < 2.78$ \gevc) indicates 
           the selected signal candidates.}
 \label{fig:trkprest_unblind} 
\end{figure}

  In the on-resonance data we find 11 events in the signal box where $5.0^{+1.8}_{-1.4}$ background events are expected. 
The probability of a background fluctuation yielding the observed number of events or more is about 4\%.
The distribution of the data in the (\DeltaE, \mes) plane is shown in figure~\ref{fig:unblind}(b). 
Figure~\ref{fig:trkprest_unblind} shows the $p*$ distribution of the on-resonance data after all other selection 
cuts have been applied. The region between the two arrows ($2.58 < p* < 2.78$ \gevc) indicates the selected signal
candidates. We set an upper limit on the \bmunu branching fraction using $\BR(\bmunu) < n_{UL}/S$ where 
$n_{UL}$ is the 90\% CL upper 
limit on the number of signal events observed and $S$ is the sensitivity of the experiment which is the product
of the signal efficiency and the number of charged \B mesons in the sample. To determine the number of 
charged \B mesons we assume equal production of \Bz and \Bp in \FourS decays. Systematic uncertainties are 
included in the limit following the prescription given in reference~\cite{Cousins}. The preliminary result is 
\begin{eqnarray}
 \BR(\bmunu) < 6.6\times 10^{-6}\nonumber
\end{eqnarray}
at the 90\% confidence level.

%%%%%%%%%%%%%%%%%%%%%%%%%%%
\section{ACKNOWLEDGMENTS}
\label{sec:Acknowledgments}
%%%%%%%%%%%%%%%%%%%%%%%%%%%

We are grateful for the 
extraordinary contributions of our \pep2\ colleagues in
achieving the excellent luminosity and machine conditions
that have made this work possible.
The success of this project also relies critically on the 
expertise and dedication of the computing organizations that 
support \babar.
The collaborating institutions wish to thank 
SLAC for its support and the kind hospitality extended to them. 
This work is supported by the
US Department of Energy
and National Science Foundation, the
Natural Sciences and Engineering Research Council (Canada),
Institute of High Energy Physics (China), the
Commissariat \`a l'Energie Atomique and
Institut National de Physique Nucl\'eaire et de Physique des Particules
(France), the
Bundesministerium f\"ur Bildung und Forschung and
Deutsche Forschungsgemeinschaft
(Germany), the
Istituto Nazionale di Fisica Nucleare (Italy),
the Foundation for Fundamental Research on Matter (The Netherlands),
the Research Council of Norway, the
Ministry of Science and Technology of the Russian Federation, and the
Particle Physics and Astronomy Research Council (United Kingdom). 
Individuals have received support from 
the A. P. Sloan Foundation, 
the Research Corporation,
and the Alexander von Humboldt Foundation.

\end{document}